\documentclass[reqno,12pt]{amsart}

\usepackage{hyperref}
\usepackage[mathscr]{eucal}
\usepackage{enumerate}
\usepackage{lineno}

\numberwithin{equation}{section}

\newtheorem{Def}{Definition}[section]
\newtheorem{Thm}[Def]{Theorem}
\newtheorem{Prp}[Def]{Proposition}
\newtheorem{Lemma}[Def]{Lemma}

\newcommand{\beq}{\begin{equation}}
\newcommand{\eeq}{\end{equation}}
\newcommand{\Proof}{\begin{proof}}
\newcommand{\QED}{\end{proof} \noindent}

\newcommand{\R}{\mathbb{R}}

\newcommand{\oneone}{C^{1,1}}
\newcommand{\oone}{C^{0,1}}

\allowdisplaybreaks[1]

\title[Regularity Singularities in General Relativity]
{No Regularity Singularities Exist at Points of General Relativistic Shock Wave Interaction between Shocks from Different Characteristic Families}

\author[M.\ Reintjes]{Moritz Reintjes}
\address{IMPA - Instituto Nacional de Matem{\'a}tica Pura e Aplicada \\ Rio de Janeiro, Brasil}
\email{moritzreintjes@gmail.com}

\author[B.\ Temple]{Blake Temple \\ \\ March 2015}
\address{Department of Mathematics\\ University of California\\ Davis, CA 95616\\ USA}
\email{temple@math.ucdavis.edu}

\begin{document}

\maketitle

\begin{abstract}
We give a constructive proof that coordinate transformations exist which raise the regularity of the gravitational metric tensor from $C^{0,1}$ to $C^{1,1}$ in a neighborhood of points of shock wave collision in General Relativity. The proof applies to collisions between shock waves coming from different characteristic families, in spherically symmetric spacetimes. Our result here implies that spacetime is locally inertial and corrects an error in our earlier RSPA-publication, which led us to the false conclusion that such coordinate transformations, which smooth the metric to $C^{1,1}$, cannot exist. Thus, our result implies that regularity singularities, (a type of mild singularity introduced in our RSPA-paper), do \emph{not exist} at points of interacting shock waves from different families in spherically symmetric spacetimes. Our result generalizes Israel's celebrated 1966 paper to the case of such shock wave interactions but our proof strategy differs fundamentally from that used by Israel and is an extension of the strategy outlined in our original RSPA-publication. Whether regularity singularities exist in more complicated shock wave solutions of the Einstein Euler equations remains open.
\end{abstract}

%

\section{Introduction}

The guiding principle in Albert Einstein's pursuit of General Relativity (GR)  was the principle that spacetime should be {\it locally inertial} \cite{Einstein}, (we say also {\it locally Minkowski} or {\it locally flat}).  That is, an observer in freefall through a gravitational field should observe all of the physics of Special Relativity, except for the second order acceleration effects due to spacetime curvature (gravity).  But the assumption that spacetime is locally inertial is equivalent to assuming the gravitational metric tensor $g$ is smooth enough so that one can pursue the construction of Riemann Normal Coordinates at a point $p$: I.e., coordinates in which $g$ is exactly the Minkowski metric at $p$, such that all first order derivatives of $g$ vanish at $p$, and all second order derivatives of $g$ are bounded in a neighborhood of $p$. However, the Einstein equations are a system of partial differential equations (PDE's) for the metric tensor $g$ coupled to the sources and the Einstein equations by themselves determine the smoothness of the gravitational metric tensor by the evolution they impose. Thus the condition on spacetime that it be locally inertial at every point cannot be assumed at the start, but must be determined by regularity theorems for the Einstein equations.

The presence of shock waves makes this issue all the more interesting for the Einstein equations with a perfect fluid source.  In this case the Einstein equations $G=\kappa T$ imply the GR compressible Euler equations $\text{Div}\, T=0$ through the Bianchi identities, \cite{Weinberg}, and the compressible Euler equations create shock waves whenever the flow is sufficiently compressive, \cite{Lax}.   At a shock wave, the fluid density, pressure, velocity, and hence  $T$ are discontinuous, so that the Einstein equations imply the curvature $G$ must also become discontinuous at shocks.  But discontinuous curvature by itself is not inconsistent with the assumption that spacetime be locally inertial.  For example, if the gravitational metric tensor were $C^{1,1}$, (differentiable with Lipschitz continuous first derivatives), then second derivatives of the metric are at worst discontinuous, and the metric has enough smoothness for there to exist coordinate transformations  which transform $g$ to the Minkowski metric at $p$, with zero derivatives at $p$, and bounded second derivatives as well, \cite{SmollerTemple}.  Furthermore, Israel's theorem asserts that a spacetime metric of regularity $C^{0,1}$, (i.e., Lipschitz continuous), across a smooth {\it single} shock surface, is lifted to $C^{1,1}$ by the $C^{1,1}$ coordinate map to Gaussian normal coordinates, and this is smooth enough to ensure the existence of locally inertial coordinate frames at each point \cite{Israel}. 

In \cite{GroahTemple}, Groah and Temple set out a framework in which to address these issues rigorously by providing the first general existence theory for spherically symmetric shock wave solutions of the Einstein-Euler equations allowing for arbitrary numbers of interacting shock waves of arbitrary strength.  In coordinates where their analysis is feasible, Standard Schwarzschild Coordinates (SSC),\footnote{A spherically symmetric metric can generically be transformed to SSC, c.f. \cite{Weinberg}}  the gravitational metric is only $C^{0,1}$ at shock waves, and it has remained an open problem as to whether the general weak solutions constructed by Groah and Temple could be smoothed to $C^{1,1}$ by coordinate transformation, as was proven by Israel for {\it single} shock surfaces, \cite{Israel}.
     
In this paper we partially resolve the open problem of Groah and Temple by proving there do exist $C^{1,1}$ coordinate transformations that lift the regularity of the gravitational metric tensor from $C^{0,1}$ to $C^{1,1}$ at a point of shock wave interaction between shocks from different characteristic families in spherically symmetric spacetimes. In \cite{ReintjesTemple} the authors introduced the idea of a {\it regularity singularity}, a point in spacetime where the metric tensor is $C^{0,1}$ but not $C^1$ regular in any coordinate system.   Our result here is the first step in extending Israel's theorem to interacting shock waves, by proving that spacetime is indeed locally inertial and that no regularity singularity exists at points of such shock wave collision.  This negates our false conclusion in \cite{ReintjesTemple} that regularity singularities exist at points of shock interaction, and the error in  \cite{ReintjesTemple} is explained and corrected in this paper, (c.f. Lemma \ref{canonicalformforJ}).   The question as to whether regularity singularities can be created in more complicated solutions of the Einstein Euler equations, by more complicated shock wave interactions, remains an open problem.

The existence of regularity singularities would be surprising for General Relativity, where it is commonly assumed that the gravitational metric tensor is at least $C^{1,1}$.   The metric regularity $C^{1,1}$  is the threshold regularity required for the existence of locally inertial coordinate frames, the existence of which is essential for properties of shock waves in Minkowski space-time to be recovered in the limit of weak gravitational fields.   The metric regularity $C^{1,1}$ is a starting assumption in the singularity theorems of Hawking and Penrose, \cite{HawkingEllis}.   At a regularity singularity the metric would be Lipschitz continuous but not $C^1$ in any coordinate system, so discontinuities in the metric derivatives would be present in every coordinate system, and this would open the door for possible new gravitational effects. The authors will address the implications of regularity singularities in a forthcoming paper.               

To state our main result precisely, let $g_{\mu\nu}$ denote a spherically symmetric spacetime metric in SSC, where the metric takes the form
\beq \label{metric SSC}
ds^2=g_{\mu\nu}dx^{\mu}dx^{\nu}=-A(t,r)dt^2+B(t,r)dr^2+r^2d\Omega^2.
\eeq
At the start either $t$ or $r$ can be taken to be timelike, and $d\Omega^2=d\vartheta^2 + \sin^2(\vartheta) d\varphi^2$ is the line element on the unit $2$-sphere, c.f. \cite{GroahTemple}.   In Section \ref{sec: point regular interaction}, we make precise the definition of a point of {\it regular shock wave interaction in SSC between shocks from different families}.  Essentially, this is a point in $(t,r)$-space where two shock waves enter and leave a point $p$, such that the metric is Lipschitz continuous across the shocks and smooth away from them, the Rankine-Hugoniont (RH) jump conditions hold across each shock curve and are continuous up to the point of interaction $p$, derivatives of all quantities are continuous up to the shock boundaries, and the SSC Einstein equations hold weakly in a neighborhood of $p$ and strongly away from the shocks, \cite{Smoller}. The main result of the paper is the following theorem:

\begin{Thm}  \label{TheoremMain} 
Suppose that $p$ is a  point of regular shock wave interaction between shocks from different families, in the sense that conditions (i) - (iv) of Definition \ref{shockinteract} hold, for an SSC metric $g_{\mu\nu}$.  Then the following are equivalent:
\begin{enumerate}[(i)]
\item There exists a $\oneone$ coordinate transformation $x^\alpha\circ(x^\mu)^{-1}$ in the $(t,r)$-plane, with Jacobian $J^\mu_\alpha$, defined in a neighborhood $\mathcal{N}$ of $p$, such that the metric components $g_{\alpha\beta} = J^\mu_\alpha J^\nu_\beta g_{\mu\nu}$ are $\oneone$ functions of the coordinates $x^{\alpha}$.
\item The Rankine Hugoniot conditions, \eqref{RHwithN1} - \eqref{RHwithN2}, hold across each shock curve in the sense of (v) of Definition \ref{shockinteract}.
\end{enumerate}
Furthermore, the above equivalence also holds for the full atlas of $C^{1,1}$ coordinate transformations, not restricted to the $(t,r)$-plane.
\end{Thm}

Our proof of Theorem \ref{TheoremMain} provides  an explicit method for constructing the Jacobians of $(t,r)$-coordinate transformations that smooth the components of the gravitational metric from $C^{0,1}$ to $C^{1,1}$ in a neighborhood of $p$. In order to prove the equivalence in Theorem \ref{TheoremMain}, we characterize all such Jacobians that lift the metric regularity.   Our method of proof differs substantially from the one used by Israel, the latter being based on studying the Einstein tensor in Gaussian normal coordinates and concluding that the metric is $C^{1,1}$ in these coordinates. The main ideas for the proof of Theorem \ref{TheoremMain} were already introduced in \cite{ReintjesTemple}, but an error in the last section led us to the wrong conclusion that metric-smoothing is not possible. In fact, Sections \ref{sec preliminaries} - \ref{Israel's Thm with new method} of this paper mostly coincide with the corresponding sections in \cite{ReintjesTemple}. In Section \ref{sec: shock_interaction}, we correct the error in \cite{ReintjesTemple} and outline the proof of Theorem \ref{TheoremMain}.

Our assumptions in Theorem \ref{TheoremMain} apply to shock wave interactions in which two timelike shock waves enter and leave the point of interaction. In the case of a perfect fluid source, this type of interaction is realized between two incoming shock waves from different characteristic families, c.f. \cite{Smoller}. 
  Although points of shock wave interaction are straightforward to construct for the relativistic compressible Euler equations in flat spacetime,  and a general existence theory for shock wave interactions in GR is given in \cite{GroahTemple}, we know of no complete mathematically rigorous construction of a point of shock wave interaction in GR sufficient to derive detailed structure at such points.  However, our assumptions regarding regular shock wave interaction in SSC given in Definition \ref{shockinteract} below, are straightforward, are consistent with \cite{GroahTemple}, and confirmed by the numerical simulations in \cite{VoglerTemple}.   

In Section \ref{sec: point regular interaction} we set out the framework of shock waves in GR, and define what we call a point of {\it regular shock wave interaction} in SSC.  In Section \ref{sec: about oone across} we introduce a canonical form for functions $\oone$ across a hypersurface and write the RH conditions as a relation between first order metric derivatives. In Section \ref{intro to method} we derive necessary and sufficient conditions that Jacobians of $C^{1,1}$ coordinate transformations lift the regularity of the metric tensor from $C^{0,1}$  to $C^{1}$ at points on a single shock surface. This is the so-called \emph{smoothing condition}. In Section \ref{Israel} we give a new constructive proof of Israel's theorem for spherically symmetric spacetimes, combining the results from Sections \ref{sec: about oone across} and \ref{intro to method}. For this, we first derive a canonical form of the Jacobian which satisfies the smoothing condition and show that the freedom to add an arbitrary $C^1$-function to our canonical form suffices for the Jacobians to be integrable to coordinate transformations. Sections 2 - 6 here agree with Sections 2 - 6 in \cite{ReintjesTemple}, in principal.           

In Section \ref{sec: shock_interaction} we extend the above strategy to the case of shock interaction and outline the proof of  Theorem \ref{TheoremMain}. The content of this section is new and complete details can be found in Section 7 - 10 of \cite{Reintjes}. For the proof, we first derive the canonical form for the Jacobian satisfying the smoothing condition across each of the shock curves. This is the source of the error in our original RSPA paper \cite{ReintjesTemple}, since we incorrectly omitted terms which encode the presence of two shocks.      As in Section \ref{Israel's Thm with new method}, we now need to prove that one can integrate the Jacobian to coordinates. This is achieved by showing that the freedom to add a $C^1$ function to the canonical form of the Jacobian suffices to solve the integrability condition, $J^\mu_{\alpha,\beta}=J^\mu_{\beta,\alpha}$, which is done as follows: choosing two of the free functions arbitrarily, (say $ \Phi^t_1$ and $\Phi^r_1$), the integrability condition turns into a linear first order system of PDE's of the form $U_t + c\, U_r = F(U)$ for the remaining two free functions $U = (\Phi^t_0,\Phi^r_0)$ as unknowns. The source term $F(U)$ is {\it non-local}, depends on the restriction of $U$ to the shock curves and derivatives of $U$ along the shocks, and is discontinuous at the shock waves. We first prove existence of a $C^{0,1}$ solution $U$, which by itself is not a sufficient smoothness for the resulting Jacobian to meet the smoothing condition. We then use a bootstrapping argument to show that $U$ is indeed $C^1$ regular if and only if the RH jump conditions hold. (Interestingly, there is an apparent loss of smoothness across the characteristic which passes through the point of shock wave interaction, even though this characteristic curve lies within the region of smoothness of the SSC metric. The resulting metric is $C^{1,1}$ but seems to be no smoother. This is a kind of mild new irregularity by itself and further indicates the subtlety of the problem.) The above construction gives us Jacobians which smooth the metric before and after the interaction takes place, but to obtain the Jacobian in a spacetime-neighborhood of the point of interaction, we need to match these two Jacobians across the surface $t=0$, the time of shock collision. Thus, in the last step we prove that one can choose the free functions at $t=0$ appropriately for the metric in the resulting coordinates to match across the $t=0$ interface such that it maintains its $C^{1,1}$ regularity. Interestingly, again the RH conditions come in at this final step to ensure that this matching can be done consistently. In this construction of the Jacobian several conditions, which appear over-determined at the start, are consistent as a consequence of the RH jump conditions, giving us confidence that all terms have now been accounted for.

\section{Preliminaries}\label{sec preliminaries}

Let $g$ denote a Lorentzian metric of signature ~$(-1,1,1,1)$ on a four dimensional spacetime manifold $M$. We call $M$ a $C^k$-manifold if it is endowed with a $C^k$-atlas, a collection of four dimensional local diffeomorphisms from $M$ to $\R^4$, such that any composition of two local diffeomorphisms $x$ and $y$ of the form ~$x\circ y^{-1}$ is $C^k$ regular. The mapping $x\circ y^{-1}$ is referred to as a coordinate transformation. In this paper we consider $\oneone$-manifolds.

Our index notation for tensors use Greek letters $\mu, \nu, \ldots\in\{t, r, \theta, \varphi\}$ for SSC coordinates (in which the spacetime metric $g$ is $C^{0,1}$) and Greek letters $\alpha, \beta, \ldots\in\{ 0,1,2,3\}$ for transformed coordinates, $g_{\alpha\beta}=\tfrac{\partial x^{\mu}}{\partial x^{\alpha}}\tfrac{\partial x^{\nu}}{\partial x^{\beta}}g_{\mu\nu}$. We use the Einstein summation convention whereby repeated up-down indices are summed over all values for the given indices. Tensors transform by contraction with the Jacobian $J^\mu_\alpha= \frac{\partial x^\mu}{\partial x^\alpha}$, the inverse Jacobian is denoted by $J^\alpha_\nu$, and indices are raised and lowered with the metric and its inverse  $g^{\mu\nu}$, which transform as bilinear forms, 
$g_{\mu\nu}= J^\alpha_\mu J^\beta_\nu g_{\alpha\beta}$. We use the fact that a matrix of functions $J^\mu_\alpha$ is the Jacobian of a regular local coordinate transformation if and only if the curls vanish, i.e.,
\beq \label{IC}
J^\mu_{\alpha,\beta} =J^\mu_{\beta,\alpha} \ \ \ \ \text{and} \ \ \ \ Det\left(J^\mu_\alpha\right) \neq 0,
\eeq
where ~$f_{,\alpha}= \frac{\partial f}{\partial x^\alpha}$ denotes partial differentiation with respect to the coordinate $x^\alpha$ and $Det\left(J^\mu_\alpha\right)$ denotes the determinant of the Jacobian, c.f. \cite{Weinberg}.

In this paper, we do not restrict to, but are motivated by the \emph{Einstein Euler equations},
\beq \label{EFE} 
G^{\mu\nu}=\kappa T^{\mu\nu},  
\eeq
which couples the metric tensor $g_{\mu\nu}$ to the undifferentiated perfect fluid sources
\beq \label{EulerStress}
 T^{\mu\nu}=(p+\rho)u^\mu u^\nu + p g^{\mu\nu},
\eeq
through the second order Einstein curvature tensor $G^{\mu\nu}= R^{\mu\nu}-\frac12 g^{\mu\nu} R$,
where
\beq \label{Euler}
\text{div}~ T=0\,
\eeq
follows from $\text{div}~G =0$.  Here ~$\kappa = 8\pi \mathcal{G}$, where $\mathcal{G}$ is Newton's gravitational constant, $\rho$ is the energy density, $u_i$ the $4$-velocity,  and $p$ the pressure, c.f. \cite{HawkingEllis}.\footnote{The Riemann curvature tensor introduced in \cite{Weinberg} differs from the one used by us and in \cite{HawkingEllis} by a factor of $-1$ which, in \cite{Weinberg}, is compensated for by setting $\kappa = -8\pi \mathcal{G}$. MAPLE uses the sign convention in \cite{Weinberg} for the Riemann tensor, which is important to keep in mind when computing the Einstein tensor for \eqref{one} - \eqref{four} with MAPLE.} Equation \eqref{Euler} reduces to the relativistic compressible Euler equations when $g_{\mu\nu}$ is the Minkowski metric, and the Euler equations close when an equation of state (e.g. $p=p(\rho)$) is imposed.  Shock waves form from smooth solutions of the relativistic compressible Euler equations when the initial data is sufficiently compressive, \cite{Smoller}.

Across a smooth shock surface $\Sigma$, the RH jump conditions hold,
\beq \label{JC}
[T^{\mu\nu}] n_\nu =0,
\eeq
where $[f]=f_L-f_R$ denotes the jump in $f$ from right to left across $\Sigma$, and $n_{\nu}$ is the surface normal.  The RH condition (\ref{JC}) is equivalent to the weak formulation of (\ref{Euler}) across $\Sigma$, c.f. \cite{Smoller}.

In this paper we restrict to time dependent spherically symmetric metrics in Standard Schwarzschild Coordinates  where the metric takes the form \eqref{metric SSC}. The Einstein equations for a metric in SSC are given by (c.f. \cite{GroahTemple})
\begin{eqnarray}\label{EFEinSSC}
B_r + B \frac{B-1}{r} &=& \kappa AB^2r T^{00} \label{one} \\
B_t &=& -\kappa AB^2r T^{01} \label{two}\\
A_r - A \frac{B-1}{r} &=& \kappa AB^2r T^{11} \label{three}\\
B_{tt} - A_{rr} + \Phi &=& -2\kappa A B r^2 T^{22} \label{four}\, ,
\end{eqnarray}
with
\[\Phi = -\frac{BA_tB_t}{2AB} - \frac{B_t^2}{2B} - \frac{A_r}{r} + \frac{AB_r}{rB} + \frac{A_r^2}{2A} + \frac{A_r B_r}{2B} \, . \]
\vspace{.1cm}
Note that the first three Einstein equations in SSC  imply that the metric cannot be any smoother than Lipschitz continuous if the source $T$ is discontinuous (for example, $T^{\mu\nu} \,\in\, L^\infty$), and in this paper we make the assumption throughout that $A$ and $B$ are Lipschitz continuous, i.e., $C^{0,1}$ functions of $t$ and $r$.

\section{A point of regular shock wave interaction in SSC between shocks from different families}\label{sec: point regular interaction}

In this paper we restrict attention to radial shock waves, by which we mean hypersurfaces $\Sigma$ locally parameterized by
\beq
\label{shock surface SSY}
\Sigma(t, \vartheta, \varphi) =(t,x(t), \vartheta, \varphi),
\eeq
across which $A$ and $B$ are $C^{0,1}$ and $T$ in (\ref{EulerStress}) satisfies (\ref{JC}).  Then, for each $t$, $\Sigma$ is a $2$-sphere with radius $x(t)$ and center $r=0$. Treating $\phi$ and $\theta$ as constant, we introduce $\gamma$, the restriction of a shock surface $\Sigma$ to the $(t,r)$-plane,
\beq \label{gamma}
\gamma(t)=(t,x(t)),
\eeq
with normal $1$-form
\beq\label{normal}
n_{\sigma}=(\dot{x},-1).
\eeq
For radial shock surfaces (\ref{shock surface SSY}) in SSC, the RH jump conditions (\ref{JC}) take the simplified form
\begin{eqnarray}
\left[T^{00}\right]\dot{x}&=&\left[T^{01}\right] \label{RHwithN1}, \\
\left[T^{10}\right]\dot{x}&=&\left[T^{11}\right] \label{RHwithN2}.
\end{eqnarray}

To generalize the above framework to collisions between shocks from different families, we think of the incoming and outgoing branches of the two shock waves as four distinct timelike shock surfaces, parameterized in SSC by
\beq \label{radialsurface1+}
\Sigma_i^{\pm}(t,\theta,\phi) =(t,x^\pm_i(t),\theta,\phi), 
\eeq
with $i=1,2$, and where $\Sigma^-_i$ is defined for $t\leq0$ and $\Sigma^+_i$ for $t\geq0$. Assume $\Sigma^\pm_i$ intersect at $t=0$, that is, 
\beq \nonumber
x^\pm_1(0)=r_0=x^\pm_2(0),
\eeq
for some $r_0>0$. Restricted to the $(t,r)$-plane, $\Sigma^\pm_i$ are described by the shock curves
\beq \label{gammacurves}
\gamma_i^\pm(t)=(t,x^\pm_i(t)),
\eeq
with normal $1$-forms
\beq\label{normali}
(n^\pm_i)_{\nu}=(\dot{x}^\pm_i,-1).
\eeq
We assume the $\gamma_i^\pm$ are $C^3$ with all derivatives extending to $t=0$. Denoting with $[\cdot]_i^\pm$ the jump across the shock curve $\gamma_i^\pm$ the RH conditions now read,
\begin{eqnarray}
\left[T^{00}\right]_i^\pm\dot{x}_i^\pm &=&\left[T^{01}\right]^\pm_i, \label{RHwithN1; two} \\
\left[T^{10}\right]^\pm_i\dot{x}_i^\pm &=&\left[T^{11}\right]^\pm_i. \label{RHwithN2; two}
\end{eqnarray}

For the proof of Theorem \ref{TheoremMain}, it suffices to restrict attention to the lower ($t<0$) or upper ($t>0$) part of a shock wave interaction that occurs at $t=0$. That is, it suffices to consider the lower or upper half plane in $\R^2$ separately, 
\beq\label{upper_lower-halfplane}
\R^2_{-}=\left\{(t,r): t < 0\right\}  \ \ \ \ \text{or} \ \ \ \R^2_{+}=\left\{(t,r): t > 0\right\},
\eeq
respectively. (We denote with $\overline{\R^2_{\pm}}$ the closure of $\R^2_\pm$.) Whenever it is clear that we restrict consideration to $\R^2_-$ or $\R^2_+$, we drop the superscript $\pm$ of the quantities introduced in \eqref{radialsurface1+} - \eqref{RHwithN2; two}.  

We now define ``a point of regular shock wave interaction in SSC between shocks from different families'' as a point $p$ where two shock waves collide, resulting in two outgoing shock waves, such that the metric is smooth away from the shock curves and Lipschitz continuous across each shock, allowing for a discontinuous $T^{\mu\nu}$ and the RH condition to hold. In the special case that $T^{\mu\nu}$ describes a perfect fluid, this type of collision corresponds to an interaction of shock waves in different characteristic families, c.f. \cite{Smoller}. 

\begin{Def}\label{shockinteract}
Let $r_0>0$, and assume $g_{\mu\nu}$ to be an SSC metric in $C^{0,1}\left(\mathcal{N} \right)$, where $\mathcal{N}\subset\R^2$ is a neighborhood of the point $p=(0,r_0)$ of intersection of the timelike shock curves $\gamma_i^\pm$, $i=1,2$, introduced in \eqref{gammacurves}. Let $\hat{\mathcal{N}}$ denote the open set consisting of all points in $\mathcal{N}$ not in the image of any $\gamma_i^\pm$. Then, we say that $p$ is a ``point of regular shock wave interaction in SSC between shocks from different families'' if:
\begin{enumerate}[(i)]
\item The pair $(g,T)$ is a strong solution of the SSC Einstein equations \eqref{one}-\eqref{four} in $\hat{\mathcal{N}}$, with $T^{\mu\nu}\in C^0(\hat{\mathcal{N}})$ and $g_{\mu\nu}\in C^2(\hat{\mathcal{N}})$.
\item The limits of $T^{\mu\nu}$ and of the metric derivatives $g_{\mu\nu,\sigma}$ exist on both sides of each shock curve $\gamma_i^\pm$, including the point $p$. 
\item The jumps in the metric derivatives $[g_{\mu\nu,\sigma}]^\pm_i(t)$ are $C^{3}$ functions for all $t\in(-\epsilon,0]$ or for all $t\,\in\,[0,\epsilon)$.    
\item The (upper/lower)-limits
\beq \nonumber
\lim\limits_{t\rightarrow0}[g_{\mu\nu,\sigma}]_i^\pm(t)=[g_{\mu\nu,\sigma}]^\pm_i(0)
\eeq
exist. The (upper/lower)-limits exist for all derivatives of $[g_{\mu\nu,\sigma}]_i^\pm$.
\item The stress tensor $T$ is bounded on $\mathcal{N}$ and satisfies the RH conditions
 \beq \nonumber
 [T^{\nu\sigma}]^\pm_i(n_i)_{\sigma}=0
 \eeq
 at each point on $\gamma^\pm_i(t)$, $t\in(-\epsilon,0)$ or $t\,\in\,(0,\epsilon)$, and the limits of these jumps exist up to $p$ as $t\rightarrow 0$.
\end{enumerate}
\end{Def}

The framework introduced above mostly agrees with our original setting in \cite{ReintjesTemple}. However, in contrast to the above definition, in \cite{ReintjesTemple} we imposed the structure only on $\R^2_-$ or $R^2_+$ separately, because there we looked for a contradiction, while here we look for a construction.

\section{Functions $\oone$ across a hypersurface}\label{sec: about oone across}

In this section we give a precise definition of functions that are $\oone$ across a hypersurface and use this to derive a canonical form for such functions.

\begin{Def}\label{Lipschitz across 1}\label{Lipschitz across; tensor}
Let  $\Sigma$ be a smooth (timelike) hypersurface in some open set $\mathcal{N} \,\subset\, \R^d$. We call a function $f$ ``Lipschitz continuous across $\Sigma$'', (or $C^{0,1}$ across $\Sigma$), if $f\,\in\, \oone(\mathcal{N})$, $f$ is smooth ($f\in C^2(\mathcal{N}\setminus\Sigma)$ suffices) in $\mathcal{N}\setminus\Sigma$, and limits of derivatives of $f$ exist and are smooth functions on each side of $\Sigma$ separately. We call a metric $g_{\mu\nu}$ Lipschitz continuous across $\Sigma$ in coordinates $x^\mu$ if all metric components are $\oone$ across $\Sigma$.
\end{Def}
The main point of the above definition is that we assume smoothness of $f$ away and tangential to the hypersurface $\Sigma$. Note that the continuity of $f$ across $\Sigma$ implies the continuity of all derivatives of $f$ tangent to $\Sigma$, i.e.,
\beq \label{Lipschitz across 1; eqn}
[f_{,\sigma}]v^{\sigma}=0,
\eeq
for all $v^{\sigma}$ tangent to $\Sigma$.  Moreover, Definition \ref{Lipschitz across 1} allows for the normal derivative of $f$ to be discontinuous, that is,
\beq\label{Lipschitz across (only), eqn}
[f_{,\sigma}]n^\sigma\neq0,
\eeq
where $n^\sigma$ is normal to $\Sigma$ with respect to some (Lorentz-) metric  $g_{\mu\nu}$ defined on $\mathcal{N}$.

We can now clarify the connections between the Einstein equations and the RH jump conditions \eqref{RHwithN1}, \eqref{RHwithN2} for SSC metrics that are only $C^{0,1}$ across a hypersurface. To this end,  consider a spherically symmetric spacetime metric (\ref{metric SSC}) given in SSC, assume that the first three Einstein equations (\ref{one})-(\ref{three}) hold, and assume that the stress tensor $T$ is discontinuous across a smooth radial shock surface described in the $(t,r)$-plane by $\gamma(t)$ as in (\ref{shock surface SSY})-(\ref{normal}).  Condition (\ref{Lipschitz across 1; eqn}) across $\gamma$ applied to each metric component $g_{\mu\nu}$ in SSC \eqref{metric SSC} then reads
\begin{eqnarray} \label{Lipschitz across; SSC}
\left[B_t\right]&=&-\dot{x} [B_r],\label{jumpone}\label{jumponeagain}\\
\left[A_t\right]&=& -\dot{x}[A_r]\label{jumptwo}\label{jumptwoagain}.
\end{eqnarray}
On the other hand, the first three Einstein equations in SSC (\ref{one})-(\ref{three}) imply
\begin{eqnarray}
 \left[B_r\right]&=&\kappa A B^2 r [T^{00}], \label{EFEinSSC1}\\
 \left[B_t\right]&=&-\kappa A B^2 r [T^{01}], \label{EFEinSSC2}\\
 \left[A_r\right]&=&\kappa A B^2 r [T^{11}].\label{EFEinSSC3}
\end{eqnarray}
Now, using the jumps in Einstein equations \eqref{EFEinSSC1}-\eqref{EFEinSSC3}, we find that \eqref{jumpone} is equivalent to the first RH jump condition \eqref{RHwithN1}, (c.f. Lemma 9, page 286, of \cite{SmollerTemple}), while the second condition (\ref{jumptwo}) is independent of equations (\ref{EFEinSSC1})-(\ref{EFEinSSC3}), because $A_t$ does not appear in the first order SSC equations (\ref{one})-(\ref{three}).
The result, then, is that in addition to the assumption that the metric be $C^{0,1}$ across the shock surface in SSC, the RH conditions (\ref{RHwithN1}) and (\ref{RHwithN2}) together with the Einstein equations (\ref{EFEinSSC1})-(\ref{EFEinSSC3}), yield only one additional condition over and above (\ref{jumpone}) and (\ref{jumptwo}), namely,
\beq
[A_r]=-\dot{x}[B_t]\; \label{[Bt]shockspeed=[Ar]}.
\eeq
The RH jump conditions together with the Einstein equations will enter our method in Sections \ref{intro to method}-\ref{sec: shock_interaction} only through the three equations (\ref{[Bt]shockspeed=[Ar]}), \eqref{jumpone} and \eqref{jumptwoagain}.

The following lemma provides a canonical form for any function $f$ that is Lipschitz continuous across a {\it single} shock curve $\gamma$ in the $(t,r)$-plane, under the assumption that the vector $n^{\mu}$, normal to $\gamma$, is obtained by raising the index in (\ref{normal}) with respect to a  Lorentzian metric $g$ that is $C^{0,1}$ across $\gamma$.  (Note that by Definition \ref{Lipschitz across; tensor}, $n^{\mu}$ varies $C^1$ in directions tangent to $\gamma$. Here, we suppress the angular coordinates.)

\begin{Lemma}\label{characerization1} Suppose $f$ is $\oone$ across a smooth curve $\gamma(t)=(t,x(t))$ in the sense of Definition \ref{Lipschitz across 1}, $t\, \in \, (-\epsilon,\epsilon)$, in an open subset $\mathcal{N}$ of $\R^2$.  Then there exists a function $\Phi \: \in \, C^1(\mathcal{N})$ such that
\beq\label{characerization1, eqn}
f(t,r)=\frac12 \varphi(t) \left|x(t)-r\right| +\Phi(t,r),
\eeq
where
\beq\label{normalcondition}
\varphi(t)=\frac{[f_{,\mu}]n^{\mu}}{n^{\sigma}n_{\sigma}}\, \in\,C^1(-\epsilon,\epsilon),
\eeq
and $n_{\mu}(t)=(\dot{x}(t),-1)$ is a $1$-form normal to the tangent vector $v^\mu(t)=\dot{\gamma}^\mu(t)$.   In particular, it suffices that indices are raised and lowered by a Lorentzian metric $g_{\mu\nu}$ which is $C^{0,1}$ across $\gamma$.
\end{Lemma}

In words, the canonical form (\ref{characerization1, eqn}) separates off the kink of $f$ across $\gamma$ (that is, the $C^{0,1}$ element of $f$) from its more regular $C^1$ behavior away from $\gamma$:  The kink is incorporated into $\left|x(t)-r\right|$,  $\varphi$ gives the smoothly varying strength of the jump, and $\Phi$ encodes the remaining $C^1$ behavior of $f$.

In Section \ref{sec: shock_interaction} we need a canonical form analogous to (\ref{characerization1, eqn}) for two shock curves, but such that it allows for the Jacobian to be in the weaker regularity class $C^{0,1}$ away from the shock curves. To this end, suppose two timelike shock surfaces described in the $(t,r)$-plane by, $\gamma_i(t)$, such that (\ref{radialsurface1+})-(\ref{RHwithN2; two}) applies. To cover the generic case of shock wave interaction, we assume each $\gamma_i(t)$ is smooth (at least $C^2$) away from $t=0$ and all derivatives extend continuously to $t=0$. It suffices to restrict to upper shock wave interactions in $\R^2_{+}$.

\begin{Lemma}\label{characerization2} Let $\gamma_i(t)=(t,x_i(t))$ be two smooth curves defined on $I=(0,\epsilon)$, for some $\epsilon>0$, such that \eqref{radialsurface1+} - \eqref{normali} hold. Let $\mathcal{N}$ be an open neighborhood of $p=(0,r_0)$ in $\R^2$ and suppose $f$ is in $\oone(\mathcal{N}\cap\R^2_+)$, but such that $f$ is $C^2$ tangential to each $\gamma_i$ with \eqref{Lipschitz across 1; eqn} holding.  Then there exists a $\oone$ function $\Phi$ defined on $\mathcal{N}\cap\R^2_{+}$,
such that
\beq\label{jumPhi} 
[\Phi_t]_i = 0 = [\Phi_r]_i,
\eeq
for $i=1,2$, and
\beq\label{canoicalfortwo}
f(t,r)= \sum_{i=1,2} \varphi_i(t) \left|x_i(t)-r\right| +\Phi(t,r),
\eeq
for all $(t,r)$ in $\mathcal{N}\cap\R^2_{+}$, where
\beq\label{kink in canonical form, two}
\varphi_i(t) = \frac12 \frac {[f_{,\mu}]_i (n_i)^{\mu}} {(n_i)^{\mu}(n_i)_\mu} \, \in C^{1}(I),
\eeq
and $(n_{i})_{\mu}(t)=(\dot{x}_i(t),-1)$ is the $1$-form normal to $v_i^\mu(t)=\dot{\gamma}_i^\mu(t)$, for $i=1,2$,  and indices are raised by a Lorentzian metric $C^{0,1}$ across each $\gamma_i$. 
\end{Lemma}

\section{A Necessary and Sufficient Condition for Smoothing Metrics}\label{intro to method}

In this section we derive a necessary and sufficient pointwise condition on the Jacobians of a coordinate transformation that it lift the regularity of a $C^{0,1}$ metric tensor to $C^{1,1}$ in a neighborhood of a point on a single shock surface $\Sigma$.  This is the starting point for our methods in Sections \ref{Israel} and \ref{sec: shock_interaction}. Proofs and further results can be found in Section 5 of \cite{Reintjes}.

We begin with the transformation law
\beq\label{metrictrans}
g_{\alpha\beta}=J^{\mu}_{\alpha} J^{\nu}_{\beta} g_{\mu\nu},
\eeq
for the metric components at a point on a hypersurface $\Sigma$ for a general $C^{1,1}$ coordinate transformation $x^{\mu}\rightarrow x^{\alpha}$, where, as customary, the indices indicate the coordinate system. $J^{\mu}_{\alpha}$ denotes the Jacobian of the transformation, that is, $J^{\mu}_{\alpha}=\frac{\partial x^{\mu}}{\partial x^{\alpha}}$.
Assume now, that the metric components $g_{\mu\nu}$ are only Lipschitz continuous with respect to $x^{\mu}$ across $\Sigma$.
Then differentiating (\ref{metrictrans}) in the direction $w=w^{\sigma}\frac{\partial}{\partial x^{\sigma}}$ we obtain
 \beq\label{basicstart}[g_{\alpha\beta,\gamma}]w^\gamma= J^\mu_\alpha J^\nu_\beta [g_{\mu\nu,\sigma}] w^\sigma + g_{\mu\nu} J^\mu_\alpha [J^\nu_{\beta,\sigma}] w^\sigma +g_{\mu\nu} J^\nu_\beta [J^\mu_{\alpha,\sigma}] w^\sigma \, ,\eeq
  where $[f]$ denotes the jump in the quantity $f$ across the shock surface $\Sigma.$   Thus, since both $g$ and $J^{\mu}_{\alpha}$ are in general Lipschitz continuous across $\Sigma$, the jumps appear only on the derivatives.  Equation (\ref{basicstart}) gives a necessary and sufficient condition for the metric $g$ to be $C^{1,1}$ in $x^{\alpha}$ coordinates. Namely,  taking $w=\frac{\partial}{\partial x^{\sigma}}$ in SSC, (\ref{basicstart}) implies that  $[g_{\alpha\beta,\gamma}]=0$ for every $\alpha, \beta,\gamma=0,...,3$ if and only if
  \beq\label{smoothingcondt1}
 [J^\mu_{\alpha,\sigma}] J^\nu_\beta g_{\mu\nu} + [J^\nu_{\beta,\sigma}]J^\mu_\alpha g_{\mu\nu} + J^\mu_\alpha J^\nu_\beta [g_{\mu\nu,\sigma}] =0 \ .
 \eeq
Note that if the coordinate transformation is $C^2$, so that $J^{\mu}_{\alpha}$ is $C^1$, then the jumps in $J$ vanish, and (\ref{basicstart}) reduces to
\beq \nonumber
[g_{\alpha\beta,\gamma}]w^\gamma= J^\mu_\alpha J^\nu_\beta [g_{\mu\nu,\sigma}] w^\sigma, 
\eeq
which is tensorial because the non-tensorial terms cancel out in the jump $[g_{\alpha\beta,\gamma}]$.   It is precisely the lack of covariance in (\ref{basicstart}) for $C^{1,1}$ transformations that provides the necessary degrees of freedom in the jumps $[J^{\mu}_{\alpha,\sigma}]$ to lift the smoothness of a Lipschitz metric one order at a single shock surface.

We now exploit linearity in (\ref{smoothingcondt1}) to solve for the $[J^\mu_{\alpha,\sigma}]$ associated with a given $C^{1,1}$ coordinate transformation.  To this end,  suppose we are given a single radial shock surface $\Sigma$ in SSC locally parameterized by
\beq \label{radialsurface} 
\Sigma(t,\theta,\phi) =(t,x(t),\theta,\phi) \, .
\eeq 
For such a hypersurface in Standard Schwarzschild Coordinates (SSC), the angular variables play a passive role, and the essential issue regarding smoothing the metric components by $C^{1,1}$ coordinate transformation, lies in the atlas of $(t,r)$-coordinate transformations.
Thus we restrict to the atlas of $(t,r)$-coordinate transformations for a general $C^{0,1}$ metric in SSC, c.f. (\ref{metric SSC}).
 The following lemma gives the unique solution $[J^\mu_{\alpha,\sigma}]$ of (\ref{smoothingcondt1}) for $(t,r)$-transformations of $C^{0,1}$ metrics $g$ in SSC.

 \begin{Lemma}\label{smoothcondition}
Let
\beq\nonumber
  g_{\mu\nu} = -A(t,r)dt^2 + B(t,r)dr^2 + r^2 d\Omega^2\, ,
\eeq
be a given metric expressed in SSC, let $\Sigma$ denote a single radial shock surface \eqref{radialsurface} across which $g$ is only Lipschitz continuous.  Then the unique solution $[J^\mu_{\alpha,\sigma}]$ of \eqref{smoothingcondt1} which satisfies the integrability condition in SSC,\footnote{We use here that  $J^\mu_{\alpha,\sigma} J^\sigma_\beta =J^\mu_{\beta,\sigma} J^\sigma_\alpha $ is equivalent to \eqref{IC}, but with derivatives taken in SSC, c.f. \cite{Reintjes} for more details.} c.f. \eqref{IC},
\beq \label{IC2}
[J^\mu_{\alpha,\sigma}] J^\sigma_\beta =[J^\mu_{\beta,\sigma}] J^\sigma_\alpha  \ , 
\eeq
is given by:
\begin{eqnarray}\label{smoothingcondt2} [J^t_{0,t}]&=&-\frac12 \left( \frac{[A_t]}{A}J^t_0 + \frac{[A_r]}{A}J^r_0  \right); \ \ \ \ \ [J^t_{0,r}]=-\frac12 \left( \frac{[A_r]}{A}J^t_0 + \frac{[B_t]}{A}J^r_0  \right) \cr [J^t_{1,t}]&=& -\frac12 \left( \frac{[A_t]}{A}J^t_1 + \frac{[A_r]}{A}J^r_1  \right); \ \ \ \ \ [J^t_{1,r}]=-\frac12 \left( \frac{[A_r]}{A}J^t_1 + \frac{[B_t]}{A}J^r_1  \right) \cr  [J^r_{0,t}]&=& -\frac12 \left( \frac{[A_r]}{B}J^t_0 + \frac{[B_t]}{B}J^r_0  \right); \ \ \ \ \ [J^r_{0,r}]= -\frac12 \left( \frac{[B_t]}{B}J^t_0 + \frac{[B_r]}{B}J^r_0  \right)   \cr [J^r_{1,t}] &=& -\frac12 \left( \frac{[A_r]}{B}J^t_1 + \frac{[B_t]}{B}J^r_1  \right); \ \ \ \ \ [J^r_{1,r}]=-\frac12 \left( \frac{[B_t]}{B}J^t_1 + \frac{[B_r]}{B}J^r_1  \right) .
\end{eqnarray}
$($We use the notation $\mu,\nu, \sigma\in\left\{t,r\right\}$ and $\alpha,\beta\in\left\{0,1\right\}$, so that $t,r$ are used to denote indices whenever they appear on the Jacobian $J$.$)$
\end{Lemma}

To avoid confusion in Section \ref{Israel's Thm with new method} and \ref{sec: shock_interaction}, we introduce the notation
\beq \label{notation_smoothing-condition}
\mathcal{J}^\mu_{\alpha\sigma} = [J^\mu_{\alpha,\sigma}]
\eeq
to denote the right hand sides in \eqref{smoothingcondt2}. 

Condition (\ref{smoothingcondt1}) is a necessary and sufficient condition for $[g_{\alpha\beta,\gamma}]=0$ at a point on a smooth single shock surface.  Because Lemma \ref{smoothcondition} tells us that we can uniquely solve (\ref{smoothingcondt1}) for the Jacobian derivatives, it follows that a necessary and sufficient condition for $[g_{\alpha\beta,\gamma}]=0$ is also that the jumps in the Jacobian derivatives be exactly the functions of the jumps in the original SSC metric components recorded in (\ref{smoothingcondt2}).   Thus, Lemma \ref{smoothcondition} implies the following lemma:

\begin{Lemma}\label{smoothingcondt equiv oneone} Let $p$ be a point on a single smooth shock curve $\gamma$, and let $g_{\mu\nu}$ be a metric tensor in SSC, which is $\oone$ across $\gamma$ in the sense of Definition \ref{Lipschitz across; tensor}. Suppose $J^\mu_\alpha$  is the Jacobian of a coordinate transformation defined on a neighborhood $\mathcal{N}$ of $p$. Then the metric in the new coordinates $g_{\alpha\beta}$ is in $\oneone(\mathcal{N})$ if and only if $J^\mu_\alpha$ satisfies \eqref{smoothingcondt2}.
\end{Lemma}

\section{Metric Smoothing on Single Shock Surfaces and a Constructive Proof of Israel's Theorem}\label{Israel's Thm with new method}\label{Israel}

In this section we outline an alternative constructive proof of Israel's Theorem for spherically symmetric spacetimes, (see Section 6 in \cite{Reintjes} for complete details).  For the proof, in light of Lemma \ref{smoothingcondt equiv oneone}, we need to construct Jacobians of coordinate transformations, defined in a neighborhood of a point on a single shock surface, that satisfy (\ref{smoothingcondt2}).  In other words, we need to introduce a set of functions, $J^{\mu}_{\alpha}$, that satisfies \eqref{smoothingcondt2} and  the integrability condition (\ref{IC}) in some neighborhood of the shock. The main theorem of this section is the following:

\begin{Thm}\label{SingleShockThm}{\rm (Israel's Theorem)}
Suppose $g_{\mu\nu}$ is an SSC metric that is $\oone$ across a radial shock surface $\Sigma$ in the sense of Definition \ref{Lipschitz across 1}, such that it solves the Einstein equations \eqref{one} - \eqref{four} strongly away from $\Sigma$ for a $T^{\mu\nu}$ which is continuous away from $\Sigma$. Let $p$ be a point on $\Sigma$. Then the following is equivalent:
\begin{enumerate}[(i)]
\item There exists a $\oneone$ coordinate transformation of the $(t,r)$-plane, defined in some neighborhood $\mathcal{N}$ of $p$, such that the transformed metric components 
are $\oneone$ functions of the new coordinates.
\item The RH conditions, \eqref{RHwithN1} - \eqref{RHwithN2}, hold on $\Sigma \cap \mathcal{N}'$ for some $\mathcal{N}' \supset \mathcal{N}$.
\end{enumerate}
Furthermore, the above equivalence also holds for the full atlas of $C^{1,1}$ coordinate transformations, not restricted to the $(t,r)$-plane.
\end{Thm}

The main step is to construct Jacobians acting on the $(t,r)$-plane that satisfy the smoothing condition \eqref{smoothingcondt2} on the shock curve, the condition that guarantees $[g_{\alpha\beta,\gamma}]=0$. The following lemma gives an explicit formula for functions $J^\mu_\alpha$ satisfying  \eqref{smoothingcondt2}.  The main point is that, in the case of single shock curves,  both the RH jump conditions and the Einstein equations are necessary and sufficient for such functions $J^\mu_\alpha$ to exist.

\begin{Lemma}\label{canonicalformforJ1}
Let $\mathcal{N}$ be a neighborhood of a point $p$, for $p$ lying on a single shock curve $\gamma$ across which the SSC metric $g_{\mu\nu}$ is Lipschitz continuous in the sense of Definition \ref{Lipschitz across 1}, and let $g_{\mu\nu}$ be defined on $\mathcal{N}$. Then, there exists functions $J^\mu_\alpha \in \oone(\mathcal{N})$ which satisfy the smoothing condition \eqref{smoothingcondt2} on $\gamma\cap\mathcal{N}$ if and only if the RH conditions \eqref{[Bt]shockspeed=[Ar]} hold on $\gamma\cap\mathcal{N}$. Furthermore, any such function $J^\mu_\alpha$ is of the ``canonical form'' 
\begin{eqnarray}\label{Jacobian1}
J^\mu_\alpha (t,r) &=& \varphi^\mu_\alpha (t) \left| x(t)-r \right| + \Phi^\mu_\alpha(t,r)
\end{eqnarray}
with 
\beq \label{Jacobian1_coeff}
\varphi^\mu_\alpha (t) =-\frac12 \mathcal{J}^\mu_{\alpha\, r}(t),
\eeq 
where $\mathcal{J}^\mu_{\alpha r}$ is defined in \eqref{notation_smoothing-condition}, $\mu \in \{t,r\}$, $\alpha\in \{0,1\}$, and ~$\Phi^\mu_\alpha\in \oone(\mathcal{N})$ satisfy
\beq\label{nojumps1}
[\partial_r \Phi^\mu_\alpha]=0=[\partial_t \Phi^\mu_\alpha].
\eeq 
Explicitly, the Jacobian coefficients are given by 
\begin{eqnarray} \label{Jacobian1_coeff_expl}
\varphi^t_0(t)&=& \frac{[A_r]\phi(t) + [B_t]\omega(t)}{4A\circ \gamma(t)} \cr
\varphi^t_1(t)&=&\frac{[A_r]\nu(t) + [B_t]\zeta(t)}{4A\circ \gamma(t)} \cr
\varphi^r_0(t)&=&\frac{[B_t]\phi(t) + [B_r]\omega(t)}{4B\circ \gamma(t)}\cr
\varphi^r_1(t)&=&\frac{[B_t]\nu(t) + [B_r]\zeta(t)}{4B\circ \gamma(t)}\, ,
\end{eqnarray}
where
\beq\label{restrictedR2fct}
\phi=\Phi^t_0\circ\gamma, \ \ \omega= \Phi^r_0\circ\gamma, \ \ \nu= \Phi^t_1\circ\gamma, \ \ \zeta= \Phi^r_1\circ\gamma\, .
\eeq
Furthermore, the above equivalence also holds for the full atlas of $C^{0,1}$ coordinate transformations, not restricted to the $(t,r)$-plane. 
\end{Lemma}

\Proof
Suppose there exists a set of $\oone$ functions $J^\mu_\alpha$ satisfying \eqref{smoothingcondt2}. It is shown in \cite{Reintjes} that these functions satisfy
\beq\label{ansatz is lipschitz across}
[J^{\mu}_{\alpha,t}]=-\dot{x}[J^{\mu}_{\alpha,r}]
\eeq
for all $\mu\in\left\{t,r\right\}$ and $\alpha\in\left\{0,1\right\}$. Combining \eqref{ansatz is lipschitz across} for the special case $\mu=t$ and $\alpha=0$ with the right hand side in \eqref{smoothingcondt2} leads to
\[
-\frac12 \left( \frac{[A_t]}{A}J^t_0 + \frac{[A_r]}{A}J^r_0  \right) = \frac{\dot{x}}{2} \left( \frac{[A_r]}{A} J^t_0 + \frac{[B_t]}{A} J^r_0  \right).
\]
Using now the jump relations for the metric tensor, \eqref{jumponeagain} - \eqref{jumptwoagain}, finally gives  $[A_r]=-\dot{x}[B_t]$, that is, the non-trivial RH condition \eqref{[Bt]shockspeed=[Ar]}.

For proving the opposite direction, as a consequence of Lemma \ref{smoothcondition}, it suffices to show that all $t$- and $r$-derivatives of the functions $J^{\mu}_{\alpha}$, defined in \eqref{Jacobian1}, satisfy (\ref{smoothingcondt2}) for all $\mu\in\left\{t,r\right\}$ and $\alpha\in\left\{0,1\right\}$. Observing that (\ref{restrictedR2fct}) implies the identities
\beq\nonumber
\phi=J^t_0\circ\gamma, \  \ \ \nu=J^t_1\circ\gamma, \  \ \ \omega=J^r_0\circ\gamma, \ \ \ \zeta=J^r_1\circ\gamma \ ,
\eeq
and using the $C^1$ matching of the functions $\Phi^\mu_\alpha$, \eqref{nojumps1}, as well as the RH conditions in the form  (\ref{jumponeagain}), (\ref{jumptwoagain}) and (\ref{[Bt]shockspeed=[Ar]}), it follows immediately that the Jacobian ansatz \eqref{Jacobian1} satisfies (\ref{smoothingcondt2}). This proves the existence of functions $J^\mu_\alpha$ satisfying the smoothing condition \eqref{smoothingcondt2}. Finally,  applying Lemma \ref{characerization1}, it follows that all functions satisfying \eqref{smoothingcondt2} assume the canonical form \eqref{Jacobian1}.

In \cite{Reintjes}, Section 6, we extend the Lemma beyond coordinate transformations in the $(t,r)$-plane and the corresponding canonical form is described.
\QED

To complete the proof of Israel's Theorem, we need to show that there exist functions $\Phi^\mu_\alpha$ such that the $J^{\mu}_{\alpha}$, defined in \eqref{Jacobian1}, satisfy the integrability condition, \eqref{IC}, that is, $J^\mu_{\alpha,\beta} = J^\mu_{\beta,\alpha} $. For this, we consider $\Phi^t_1$ and $\Phi^r_1$ as given $C^2$ functions and write \eqref{IC} as a PDE in the unknown 
\beq\nonumber
U = (\Phi^t_0, \Phi^r_0),
\eeq
and it is straightforward to show that \eqref{IC} is equivalent to the system of PDE's
\beq \label{Israel_IC_PDE}
\partial_t U + c\ \partial_r U \, -\, \mathcal{M}\, U  =\, \Big(  |X| \mathcal{M}  - H(X) \left(\dot{x} - c \right)  
\Big) \left(\begin{array}{c} \varphi^t_{0} \cr \varphi^r_{0} \end{array} \right)   
- |X| \left(\begin{array}{c} \dot{\varphi}^t_{0} \cr \dot{\varphi}^r_{0} \end{array} \right)  ,
\eeq
where $X(t,r)=x(t)-r$ and the coefficients are given by
\beq \nonumber 
c = \frac{J^r_1}{J^t_1} \ \ \ \ \ \text{and} \ \ \ \ \
\mathcal{M} =   \frac{1}{J^t_1} \left( \begin{array}{cc} J^t_{1,t}  &  J^t_{1,r} \cr J^r_{1,t} & J^r_{1,r}   \end{array} \right).
\eeq
The goal now is to prove we can solve \eqref{Israel_IC_PDE} for $U \in C^1(\mathcal{N})\cap C^{2}(\mathcal{N}\setminus \gamma)$. 

Equation \eqref{Israel_IC_PDE} is a system of \emph{non-local} PDE's, since the right hand side of \eqref{Israel_IC_PDE} contains the Jacobian coefficients $\varphi^t_0$ and $\varphi^r_0$ which depend on $U\circ \gamma$ itself, and standard existence theory cannot be applied right away. Nevertheless, prescribing initial data on the shock curve, the right hand side of \eqref{Israel_IC_PDE} becomes a given source term and \eqref{Israel_IC_PDE} turns into a strictly hyperbolic linear system of first order PDE's. Imposing the condition
\beq \label{shock not characteristic}
\zeta\neq \dot{x} \nu,
\eeq
which ensures that the shock curve is non-characteristic, the standard existence theory in \cite{John} proves existence of a solution $U$ of \eqref{Israel_IC_PDE}, by integration of the initial data and the source term along characteristic lines, c.f. \cite{Reintjes}, Section 6 for more details. The existence theory yields a solution $U$ which lies in $C^{0,1}(\mathcal{N}) \cap C^{2}(\mathcal{N}\setminus \gamma)$ and is smooth away from $\gamma$, but it does not give us the necessary $C^1$ regularity across the shock, \eqref{nojumps1}, due to the presence of the (discontinuous) Heaviside functions $H(X)$ in \eqref{Israel_IC_PDE}. The final step to complete the proof of Israel's Theorem is now to show that the coefficients of $H(X)$ in \eqref{Israel_IC_PDE} vanish on the shock curve precisely when the RH jump conditions hold, as stated in the next lemma, which then yields the desired $C^1$ regularity across $\gamma$.

\begin{Lemma} \label{techlemma2}
Assume the assumptions of Theorem \ref{SingleShockThm} and denote with $f$ and $h$ the coefficient functions of the Heaviside function $H(X)$ in the first and second component of \eqref{Israel_IC_PDE}, respectively. Then,
\beq\label{f=0=h}
f\circ\gamma = 0 = h\circ\gamma
\eeq
if and only if the RH conditions, \eqref{RHwithN1} - \eqref{RHwithN2}, hold on $\gamma$. 
\end{Lemma}
\Proof
To derive an explicit expression for the coefficients to $H(X)$ in \eqref{Israel_IC_PDE}, note that the matrix $\mathcal{M}$ contains Heaviside functions as well. Then, collecting all terms containing $H(X)$ and using  $X\circ\gamma=0$ and \eqref{restrictedR2fct}, we find 
\begin{eqnarray} \label{f and h on gamma}
f\circ \gamma &=& \varphi^t_0 \, \dot{x}\, \nu -\varphi^t_1\, \dot{x}\, \phi +\varphi^t_1\, \omega -\varphi^t_0 \,\zeta \, ,\cr
h\circ \gamma &=& \varphi^r_0\, \dot{x}\, \nu - \varphi^r_1\, \dot{x}\, \phi +\varphi^r_1\, \omega - \varphi^r_0 \, \zeta.
\end{eqnarray}
Now, replace $\varphi^t_0$ and $\varphi^t_1$ by their definition, \eqref{Jacobian1_coeff_expl}, then a straightforward computation shows that $f\circ \gamma = 0$
is equivalent to
\beq \label{techlemma2, eqn1}
\left( [A_r] + \dot{x} [B_t] \right) \left( \phi \zeta - \nu \omega \right) = 0.
\eeq
Now, using
\beq \label{techlemma2, eqn2}
\left( \phi \zeta - \nu \omega \right) = \det \left( J^\mu_\alpha \circ \gamma \right) \neq 0,
\eeq
we conclude that $f\circ \gamma =0$ if and only if $[A_r] + \dot{x} [B_t] =0$, which is equivalent to the RH condition \eqref{RHwithN2}, c.f.  \eqref{[Bt]shockspeed=[Ar]}. Similarly, replacing $\varphi^r_0$ and $\varphi^r_1$ in \eqref{f and h on gamma} by \eqref{Jacobian1_coeff_expl}, a straightforward computation shows that $h\circ \gamma =0$ is equivalent to 
\beq \label{techlemma2, eqn3}
\left( [B_t] + \dot{x} [B_r] \right) \left( \phi \zeta - \nu \omega \right) =0.
\eeq
Now, using again \eqref{techlemma2, eqn2}, the equivalence of $h\circ \gamma =0$ and \eqref{jumptwo} follows, and thus to the RH condition \eqref{RHwithN1}. This completes the proof.
\QED

We can now complete the proof of Israel's Theorem.  As shown above, there exist functions $\Phi^\mu_\alpha$ such that the $J^\mu_\alpha$ defined in \eqref{Jacobian1} solve the integrability condition, \eqref{IC}. Moreover, by Lemma \ref{techlemma2}, these $\Phi^\mu_\alpha$ have the required regularity, that is, they satisfy  \eqref{nojumps1} at the shock curve and $C^2$ away from the shocks. By Lemma \eqref{canonicalformforJ1}, the $J^{\mu}_{\alpha}$ satisfy the smoothing condition (\ref{smoothingcondt2}) if and only if the RH jump conditions hold. Taken all together, we constructed a Jacobian $J^{\mu}_{\alpha}$ which is integrable to coordinate functions and which maps the Lipschitz continuous SSC metric to a metric that is $C^{1,1}$regular in the resulting coordinates if and only if the RH jump conditions hold. (See \cite{Reintjes}, Section 6, for the complete proof.) This proves Theorem \ref{Israel}.

\section{Metric Smoothing around Points of Shock Wave Interaction}\label{sec: shock_interaction}

In this section, we outline the proof of Theorem \ref{TheoremMain}, the details of which can be found in \cite{Reintjes}, Section 7 - 10. In principal, we follow the ideas from the previous section: We first extend our Jacobian ansatz \eqref{Jacobian1} to the case of two interacting shock waves and then show that this set of functions can be integrated to coordinates. However, in contrast to the single shock case addressed in the previous section, we have to pursue the construction of $J^\mu_\alpha$ on the upper and lower half-plane, $\R^2_\pm$, separately, and then show that the resulting functions can be ``glued'' together in a way appropriate to smooth the metric.     

We now begin constructing the Jacobian. In contrast to the single shock case, the restriction $J^\mu_\alpha \circ \gamma_i$, $i=1,2$, does not only depend on the free functions $\Phi^\mu_\alpha$ but also on the Jacobian coefficients from the other shock curve, $j\neq i$, that is, 
\beq \label{errorsource}
J^\mu_\alpha \circ \gamma_i = (\varphi_j)^\mu_{\alpha}\, \big|x_i(\cdot) - x_j(\cdot)\big| + \Phi^\mu_{\alpha} \circ \gamma_i, \ \ \ \ \ \ \  j\neq i.
\eeq
The error in \cite{ReintjesTemple} was to falsely omit the term $(\varphi_j)^\mu_{\alpha}\, \big|x_1(\cdot) - x_2(\cdot)\big|$ in \eqref{errorsource}. Since the smoothing conditions \eqref{smoothingcondt2} depend on $J^\mu_\alpha \circ \gamma_i$ itself, and since $(\varphi_2)^\mu_\alpha$ depends on $(\varphi_1)^\mu_\alpha$ and $(\varphi_1)^\mu_\alpha$ on $(\varphi_2)^\mu_\alpha$, we have to prove that the Jacobian coefficients, $(\varphi_i)^\mu_{\alpha}$, are well defined, in the sense that they can be consistently defined in terms of the metric and the free functions $\Phi^\mu_\alpha$ alone. The following proposition is the key step in extending Israel's Theorem to shock interactions. It gives the canonical form of Jacobians that meet the smoothing conditions across each shock curve in either $\R^2_+$ or $\R^2_-$, and act on the $(t,r)$-plane only. Without loss of generality we formulate the proposition for $\R^2_+$.

\begin{Prp}\label{canonicalformforJ}
Let $p$ be a point of regular shock wave interaction in SSC between shocks from different families, in the sense of Definition \ref{shockinteract} with (i) - (iv) being met, with corresponding SSC metric, $g_{\mu\nu}$, defined on $\mathcal{N}\cap \overline{\R^2_+}$. Then the following are equivalent:
\begin{enumerate}[(i)]
\item There exists functions $J^\mu_\alpha \in \oone \left(\mathcal{N}\cap \overline{\R^2_+} \right)$, for $\mu \in \{t,r\}$ and $\alpha \in \{0,1\}$, which satisfy the smoothing condition \eqref{smoothingcondt2} on $\gamma_i\cap\mathcal{N} \cap \overline{\R^2_+}$, for $i=1,2$.
\item The RH condition \eqref{[Bt]shockspeed=[Ar]} holds on each shock curve $\gamma_i\cap\mathcal{N} \cap \overline{\R^2_+}$, for $i=1,2$, as in Definition \ref{shockinteract}, (v).
\end{enumerate}
Furthermore, any such set of functions $J^\mu_\alpha$ is of the ``canonical form'' 
\beq\label{Jacobian2}
J^\mu_\alpha(t,r)= \sum_{i=1,2} (\varphi_i)^\mu_{\alpha}(t) \left| x_i(t)-r \right| + \Phi^\mu_\alpha(t,r),
\eeq
where $\Phi^\mu_\alpha \in \oone \left(\mathcal{N}\cap \overline{\R^2_+} \right)$ have matching derivatives across each shock curve $\gamma_i(t)$, for $t>0$, that is,
\beq\label{nojumps2}
[\partial_r \Phi^\mu_{\alpha}]_i \ = \ 0 \ = \ [\partial_t \Phi^\mu_{\alpha}]_i  \ \ \ \ \forall \, \mu \, \in \, \{t,r\},  \ \forall \, \alpha \, \in\, \{0,1\},
\eeq
and where $(\varphi_i)^\mu_\alpha$ is defined implicitly through
\beq\label{Jacobiancoeff2_impl}
(\varphi_i)^\mu_\alpha =  -\frac12 (\mathcal{J}_i)^\mu_{\alpha r}  
\eeq
with $(\mathcal{J}_i)^\mu_{\alpha r}$ denoting the values $\mathcal{J}^\mu_{\alpha r}$ in \eqref{smoothingcondt2} with respect to $\gamma_i$. Explicitly, the values for $(\varphi_i)^\mu_\alpha$ are given by
\begin{eqnarray}
(\varphi_i)^t_{0} = -\frac{B_i}{A_i} \, \dot{x}_i \, (\varphi_i)^r_{0}, \label{Jacobiancoeff2_expl_t-0} \\
(\varphi_i)^t_{1} = -\frac{B_i}{A_i}\, \dot{x}_i \, (\varphi_i)^r_{1}, \label{Jacobiancoeff2_expl_t-1} 
\end{eqnarray}
\begin{align}
(\varphi_i)^r_{0}   =  
\frac{ \frac1{4B_i} \Big(  [B_t]_i \, \Phi^t_{0}|_i   +   [B_r]_i \, \Phi^r_{0}|_i   \Big)  +  \frac{1}{4 B_j}  \Big(  [B_t]_j \, \Phi^t_{0}|_j   +   [B_r]_j\,   \Phi^r_{0}|_j   \Big) \mathcal{B}_{ij} }      {1 -   \mathcal{B}_{ij} \mathcal{B}_{ji} } \, , \label{Jacobiancoeff2_expl_r-0} \\
(\varphi_i)^r_{1} = 
\frac{\frac1{4B_i} \Big(  [B_t]_i \, \Phi^t_{1}|_i  +   [B_r]_i \, \Phi^r_{1}|_i   \Big)  +  \frac{1}{4 B_j}  \Big(  [B_t]_j \, \Phi^t_{1}|_j + [B_r]_j\,   \Phi^r_{1}|_j   \Big) \mathcal{B}_{ij} }       {1 -   \mathcal{B}_{ij} \mathcal{B}_{ji} } \, , \label{Jacobiancoeff2_expl_r-1} 
\end{align}
with $j\neq i$ in \eqref{Jacobiancoeff2_expl_r-0} and \eqref{Jacobiancoeff2_expl_r-1}, and where we define $A_i = A \circ \gamma_i$, $B_i = B \circ \gamma_i$,
\beq \label{restrictedR2fct, interaction}
\Phi^\mu_{\alpha}|_i= \Phi^\mu_\alpha \circ \gamma_i \, 
\eeq
and 
\beq\label{Jacobiancoeff2_expl_B}
\mathcal{B}_{ij} = \frac{|x_1(\cdot)-x_2(\cdot)|}{4B_i}  \left(  [B_r]_i   - \frac{B_j}{A_j} \dot{x}_j \, [B_t]_i  \right).
\eeq 
Furthermore, the above equivalence also holds for the full atlas of $C^{1,1}$ coordinate transformations, not restricted to the $(t,r)$-plane.
\end{Prp}

The proof of Proposition \ref{canonicalformforJ} is recorded in \cite{Reintjes}, Section 7. To reiterate, the error in \cite{ReintjesTemple}, was to falsely omit the term $(\varphi_j)^\mu_{\alpha}\, \big|x_1(\cdot) - x_2(\cdot)\big|$ in \eqref{errorsource}, which we correct here. The effect on the Jacobian coefficients \eqref{Jacobiancoeff2_expl_r-0} - \eqref{Jacobiancoeff2_expl_r-1} is precisely the appearance of the non-zero function $\mathcal{B}_{ij}$, and \eqref{Jacobiancoeff2_expl_r-0} - \eqref{Jacobiancoeff2_expl_r-1} reduce to the (incorrect) formulas in \cite{ReintjesTemple} upon setting $\mathcal{B}_{ij} = 0$.

Following the argument in Section \ref{Israel's Thm with new method}, the next step in the construction is to first write the integrability condition as a PDE in the unknown $U= \, ^T(\Phi^t_0, \Phi^r_0)$, considering $\Phi^t_1$ and $\Phi^r_1$ as given smooth functions which enter the coefficients of the PDE, and then to prove existence of a suitable regular solution $U$. As in Section \ref{Israel's Thm with new method}, we write the integrability conditions as
\beq \label{IC_2shocks}
\partial_t U + c\, \partial_r U \,=\, F(U),
\eeq
where 
\begin{eqnarray} \nonumber 
c = \frac{J^r_1}{J^t_1}, \ \ \ \ \  
\mathcal{M} =   \frac{1}{J^t_1} \left( \begin{array}{cc} J^t_{1,t}  &  J^t_{1,r} \cr J^r_{1,t} & J^r_{1,r}   \end{array} \right)
\end{eqnarray}
and, setting $X_i(t,r)=x_i(t)-r$, 
\beq \nonumber 
F = \mathcal{M}\, U    +    \sum_{i=1,2} \left\{ \Big(  |X_i| \mathcal{M}  - H(X_i) \left(\dot{x}_i - c \right)  
\Big) \left(\begin{array}{c} (\varphi_i)^t_{0} \cr (\varphi_i)^r_{0} \end{array} \right)   
- |X_i| \left(\begin{array}{c} (\dot{\varphi}_i)^t_{0} \cr (\dot{\varphi}_i)^r_{0} \end{array} \right)    \right\}.
\eeq

Again, the difficulty proving the existence of solutions to \eqref{IC_2shocks} is that $F(U)$ contains the \emph{non-local} source terms $(\varphi_i)^\mu_{0}$ and $(\dot{\varphi}_i)^t_{0}$, which depend on $U\circ \gamma_i$ and its derivatives. In contrast to the single shock case, localizing the $F(U)$ by imposing initial data on the shock curves is problematic because of the lack of regularity at the point of interaction and the subsequent gluing problem. For this reason, we develop an iterative scheme in which, we replace $F(U^k)$ by $F(U^{k-1})$, where $U^{k-1}$ is the known prior iterate at the $k$-th step. One of the problems proving convergence of this scheme is controlling the derivatives in $F(U^{k-1})$, which are of leading order in \eqref{IC_2shocks}. These terms can be controlled, since all these derivatives are multiplied by $|x_i(t)-r|$, $i=1,2$, which are small close to the point of interaction. In Section 8 of \cite{Reintjes}, we prove that these iterates indeed converge to a Lipschitz solution of \eqref{IC_2shocks}. In generalization of Lemma \ref{techlemma2}, it is then shown that this solution has the crucial $C^1$ regularity across the shocks, \eqref{nojumps2}, necessary for the construction of the Jacobian smoothing the metric tensor from $\oone$ to $C^{1,1}$, c.f. Proposition \ref{canonicalformforJ}.

In Section 8 of \cite{Reintjes} we prove that these iterates indeed converge to a solution of \eqref{IC_2shocks} with the required $C^1$ regularity. This is accomplished by first proving the existence of a Lipschitz continuous solution and then bootstrapping to the crucial $C^1$ regularity. This bootstrapping requires the RH jump conditions, as in the proof of Lemma \ref{techlemma2}, but a further difficulty is that the regularity along the charactersitic curve emanating from the point of shock interaction must also be addressed.
See Section 8 in \cite{Reintjes} for details. The result is recorded in the following proposition, we formulate it in $\R^2_+$, but the same result holds in $\R^2_-$.

\begin{Prp} \label{soln_IC_Prop}
Assume $C^2$ regular initial data $U_0(r)$ and assume $\Phi^t_1$ and $\Phi^r_1$ are given $C^3$ functions. Then, there exist a neighborhood $\mathcal{N}$ of $p$ and there exist a $C^{1,1}$ regular function $U= \, ^T(\Phi^t_0, \Phi^r_0)$ which solves the integrability condition, \eqref{IC_2shocks}, in the region $\mathcal{N}_+ = \mathcal{N}\cap\overline{\mathbb{R}^2_+}$, such that $U(0,r)=U_0(r)$ for all $(0,r)\in\mathcal{N}_+$.
\end{Prp}

Proposition \ref{soln_IC_Prop} finalizes the construction of Jacobians on $\R^2_+$, which are integrable to coordinate transformations and map the Lipschitz continuous SSC metric to a $C^{1,1}$ regular metric in the new coordinates. A similar construction gives Jacobians with the same properties on $\R^2_-$. To complete the proof of Theorem \ref{TheoremMain}, it remains only to prove that one can ``glue'' these Jacobians at the $(t=0)$-interface and maintain the $C^{1,1}$ metric regularity, by a suitable choice of the free functions $\Phi^\mu_\alpha$ at $t=0$.

For this, we first introduce some notation. We denote all objects in \eqref{Jacobian2} - \eqref{Jacobiancoeff2_expl_B} with an additional index ``$+$'' or ``$-$'' to indicate whether they are defined in $\overline{\R^2_+}$ or $\overline{\R^2_-}$, respectively. For instance, $J^{\mu\pm}_\alpha $ 
denotes the canonical Jacobian on $\mathcal{N}_\pm =\mathcal{N}\cap \overline{\R^2_\pm}$, $(\varphi_i^\pm)^\mu_{\alpha}$ its coefficients and $\Phi^{\mu\pm}_\alpha$ its free functions. We denote with $\{\cdot\}$ the jump across the $(t=0)$-interface, that is,  
\begin{eqnarray} \label{def_curly_brackets}
\{u \}(r) &=& \lim_{t\nearrow0} u(t,r) - \lim_{t\searrow0} u(t,r) , \ \ \  \ \text{for} \ r\neq r_0, \cr
\{u\}(r_0) &=& u_{M^-} - \, u_{M^+} ,
\end{eqnarray}
where $u$ is some function for which the above limits are well-defined, e.g., the metric or the Jacobian derivatives, and where $u_{M^-}$ denotes the limit at $p$ of $u$ restricted to the region in $\R^2_-$ between the two shock curves and $u_{M^+}$ the respective limit between the upper two shock curves. 

We now derive the conditions for matching $J^{\mu\pm}_{\alpha}$, conditions which are necessary and sufficient for the metric in the new coordinates, $g_{\alpha\beta}=J^\mu_\alpha J^\nu_{\beta} g_{\mu\nu}$, to be $C^{1,1}$ regular across the $(t=0)$-interface. The condition that the Jacobian matches continuously across the $(t=0)$-interface is 
\beq\label{matching_C0}
\{ J^\mu_\alpha  \}(r) = J^{\mu-}_{\alpha}(0,r) - J^{\mu+}_{\alpha}(0,r)  \, = 0, 
\eeq
for all $r$. We call these the \emph{$C^0$-matching conditions}, c.f. \cite{Reintjes}.

For the matching of the Jacobian derivatives, we follow the reasoning in Section \ref{intro to method} which leads to the smoothing condition, \eqref{smoothingcondt1}, but now we apply this reasoning to the $(t=0)$-interface. That is, the condition that $g_{\alpha\beta}$ is continuously differentiable across the $(t=0)$-interface is given by 
\beq\label{matching_metric}
\{ g_{\alpha\beta,\sigma} \} = 0.
\eeq
Substituting $g_{\alpha\beta} = J^\mu_\alpha J^\nu_\beta g_{\mu\nu}$ into \eqref{matching_metric} and using \eqref{matching_C0} as well as the SSC-metric being $C^1$ regular away from the shocks, i.e., $\{g_{\mu\nu,\sigma}\}(r)=0$ for all $r\neq r_0$, we conclude that the \emph{$C^1$-matching conditions} are given by
\begin{eqnarray}
\left( \{J^\mu_{\alpha,\sigma}\} J^\nu_\beta + \{J^\nu_{\beta,\sigma}\} J^\mu_\alpha \right) g_{\mu\nu} &=& 0, \ \ \ \ \ \ \ \ \ \ \ \ \ \ \ \ \ \ \ \ \  \forall \, r\neq r_0,  \label{matching_C1_r=r}  \\
\left( \{J^\mu_{\alpha,\sigma}\} J^\nu_\beta + \{J^\nu_{\beta,\sigma}\} J^\mu_\alpha \right) g_{\mu\nu} &=& -J^\mu_\alpha J^\nu_\beta \{g_{\mu\nu,\sigma}\}, \ \ \ \ \ \text{at} \ r=r_0.  \label{matching_C1_r=r0}
\end{eqnarray}

To finish the proof of Theorem \ref{TheoremMain}, we now show that \eqref{matching_C0} and \eqref{matching_C1_r=r} - \eqref{matching_C1_r=r0} are met for the Jacobian in \eqref{Jacobian2}, by appropriately matching the free functions $\Phi^{\mu+}_\alpha$ and $\Phi^{\mu-}_\alpha$ as well as their $t$-derivatives at $t=0$. At the start, the conditions \eqref{matching_C0} and \eqref{matching_C1_r=r} - \eqref{matching_C1_r=r0} appear over-determined, essentially because the derivatives  $\partial_t U^\pm=(\partial_t\Phi^{t\pm}_{0}, \partial_t\Phi^{r\pm}_{0})$ are not free to assign, but determined by equation \eqref{IC_2shocks} at $t=0$.  Nevertheless, \eqref{IC_2shocks} together with the RH jump conditions give the consistency of the matching conditions, \eqref{matching_C0} and \eqref{matching_C1_r=r} - \eqref{matching_C1_r=r0}, within the freedom available and imply exactly the matching of $\Phi^{\mu+}_\alpha$ and $\Phi^{\mu-}_\alpha$. This is all achieved in the final lemma, the proof of which is recorded in \cite{Reintjes}, Section 9.

\begin{Lemma}\label{matching_lemma}
Let $J^{\mu\pm}_\alpha$ be two Jacobians of the canonical form \eqref{Jacobian2}, defined on $\mathcal{N}_\pm = \mathcal{N}\cap \overline{\R^2_\pm}$ respectively, with corresponding free functions $\Phi^{\mu\pm}_\alpha$. Assume that the integrability condition \eqref{IC} holds and that $J^t_1(0,r)\neq0$, by an appropriate choice of $\Phi^t_1(0,r)$. If the $\Phi^{\mu\pm}_\alpha$ match at $t=0$, such that 
\begin{eqnarray}
\{\Phi^\mu_\alpha\}(r) &=& 0, \label{matching_C0_a} \\
\{\partial_t \Phi^t_{1} \} (r) &=& 
-\Big( (\dot{\varphi}_1^-)^{t}_{1} + (\dot{\varphi}_2^-)^{t}_{1} - (\dot{\varphi}_1^+)^{t}_{1} - (\dot{\varphi}_2^+)^{t}_{1}  \Big) |r-r_0|,  \label{matching_C1_t_first}  \\ 
\{\partial_t\Phi^r_{1} \} (r) &=& 
-\Big( (\dot{\varphi}_1^-)^{r}_{1} + (\dot{\varphi}_2^-)^{r}_{1} - (\dot{\varphi}_1^+)^{r}_{1} - (\dot{\varphi}_2^+)^{r}_{1}  \Big) |r-r_0| \label{matching_C1_t_second}
\end{eqnarray}
hold for all  $(0,r)\in \mathcal{N}$, then $J^{\mu\pm}_\alpha$ satisfies \eqref{matching_C0} and \eqref{matching_C1_r=r} - \eqref{matching_C1_r=r0}.
\end{Lemma}

Since we are free to choose the initial data of \eqref{IC_2shocks} as well as the remaining free functions, $\Phi^t_{1}$ and $\Phi^r_{1}$, such that \eqref{matching_C0_a} - \eqref{matching_C1_t_second} and $J^t_1(0,r)\neq0$ hold, Lemma \ref{matching_lemma} implies that our canonical Jacobian can be matched such that the metric in the new coordinates is $C^{1,1}$ regular. This completes the proof of Theorem \ref{TheoremMain}.

\section{Conclusion}

Our result shows that no regularity singularities exist at  points of shock wave interaction between shocks from different characteristic families, and this corrects the false conclusion in \cite{ReintjesTemple}. We prove that one can extend Israel's result to shock wave solutions of the Einstein equations containing points of shock wave interaction between shocks from different characteristic families in SSC. We introduce a new method for constructing Jacobians of coordinate transformations that map the $C^{0,1}$ SSC metric to a $C^{1,1}$ regular metric in the new coordinates. Our method differs fundamentally from Israel's proof based on Gaussian normal coordinates.   Israel's proof does not extend because Gaussian normal coordinates do not exist in a neighborhood of a point of shock wave interaction.    Whether regularity singularities can be created by shock interactions  more complicated than the interaction of two spherical shock waves in SSC from different families, remains an open problem, even assuming spherical symmetry.   But our method opens the door to address this regularity issue for more complicated solutions.  Perfect fluid matter models are essential to the description of astrophysical phenomena.    Thus the question as to the existence of regularity singularities is fundamental to General Relativity, both because their existence would change the mathematical framework for  GR perfect fluids, and because they would give rise to new detectable astrophysical effects.

%

\section*{Authors Contributions}

The ideas and methods presented here are the creation of M. Reintjes. The detailed proofs are due to him and can be found in \cite{Reintjes}. Many of these ideas were first introduced in his dissertation, which was supervised by B. Temple. 

\section*{Acknowledgments}

M.Reintjes is grateful to the Department of Mathematics of the University of Michigan - Ann Arbor for hosting him from January 2013 until August 2013, with special thanks to Lydia Bieri and Joel Smoller. Moreover, M. Reintjes is thankful to IMPA for hosting him from September 2013 until December 2014, in particular to Dan Marchesin and Hermano Frid.

\section*{Funding}

M. Reintjes was supported by the Deutsche Forschungsgemeinschaft (DFG), Grant Number RE 3471/2-1, from January 2013 until December 2014. Since January 2015 M. Reintjes is a Post-Doctorate at IMPA, funded through CAPES-Brazil. 
B. Temple was supported by NSF Applied Mathematics Grant Number DMS-010-2493.

\providecommand{\bysame}{\leavevmode\hbox to3em{\hrulefill}\thinspace}
\providecommand{\MR}{\relax\ifhmode\unskip\space\fi MR }
\providecommand{\MRhref}[2]{%
  \href{http://www.ams.org/mathscinet-getitem?mr=#1}{#2}
}
\providecommand{\href}[2]{#2}

\end{document}